\documentclass[aps,twocolumn,prd,showpacs]{revtex4}
\usepackage{epsfig}
\usepackage{bm}
\usepackage{dcolumn}
\usepackage{multirow}

\def\reff@jnl#1{{\rm#1\/}}
\def\bi#1{\hbox{\boldmath{$#1$}}}

\def\aj{\reff@jnl{AJ}}                  
\def\araa{\reff@jnl{ARA\&A}}            
\def\apj{\reff@jnl{ApJ}}                        
\def\apjl{\reff@jnl{ApJ}}               
\def\apjs{\reff@jnl{ApJS}}              
\def\ao{\reff@jnl{Appl.Optics}}         
\def\apss{\reff@jnl{Ap\&SS}}            
\def\aap{\reff@jnl{A\&A}}               
\def\aapr{\reff@jnl{A\&A~Rev.}}         
\def\aaps{\reff@jnl{A\&AS}}             
\def\azh{\reff@jnl{AZh}}                        
\def\baas{\reff@jnl{BAAS}}              
\def\jrasc{\reff@jnl{JRASC}}            
\def\memras{\reff@jnl{MmRAS}}           
\def\mnras{\reff@jnl{MNRAS}}            
\def\pra{\reff@jnl{Phys.Rev.A}}         
\def\prb{\reff@jnl{Phys.Rev.B}}         
\def\prc{\reff@jnl{Phys.Rev.C}}         
\def\prd{\reff@jnl{Phys.Rev.D}}         
\def\prl{\reff@jnl{Phys.Rev.Lett}}      
\def\pasp{\reff@jnl{PASP}}              
\def\pasj{\reff@jnl{PASJ}}              
\def\qjras{\reff@jnl{QJRAS}}            
\def\skytel{\reff@jnl{S\&T}}            
\def\solphys{\reff@jnl{Solar~Phys.}}    
\def\sovast{\reff@jnl{Soviet~Ast.}}     
\def\ssr{\reff@jnl{Space~Sci.Rev.}}     
\def\zap{\reff@jnl{ZAp}}                        
\def\nat{\reff@jnl{Nature}}             
\newcommand{\CC}{\mathsf{C}}

\begin{document}
\title{Exact likelihood evaluations and foreground marginalization in
low resolution WMAP data} 
\author{An\v ze Slosar} 
\affiliation{Faculty of Mathemathics and Physics, University of Ljubljana, Slovenia}
\author{Uro\v{s} Seljak}
\author{Alexey Makarov}
\affiliation{Department of Physics, Princeton University, Princeton NJ 08544, U.S.A.}

\date{\today}

\begin{abstract}

The large scale anisotropies of WMAP data have attracted a lot of
attention and have been a source of controversy, with many of
favourite cosmological models being apparently disfavoured by the
power spectrum estimates at low $\ell$.  All of the existing analyses
of theoretical models are based on approximations for the likelihood
function, which are likely to be inaccurate on large scales.  Here we
present exact evaluations of the likelihood of the low multipoles by
direct inversion of the theoretical covariance matrix for low
resolution WMAP maps.  We project out the unwanted galactic
contaminants using the WMAP derived maps of these foregrounds. This
improves over the template based foreground subtraction used in the
original analysis, which can remove some of the cosmological signal
and may lead to a suppression of power. As a result we find an
increase in power at low multipoles.  For the quadrupole the maximum
likelihood values are rather uncertain and vary between 140-220$\mu
{\rm K}^2$. On the other hand, the probability distribution away from
the peak is robust and, assuming a uniform prior between 0 and
$2000\mu {\rm K}^2$, the probability of having the true value above
$1200\mu {\rm K}^2$ (as predicted by the simplest $\Lambda CDM$ model)
is 10\%, a factor of 2.5 higher than predicted by WMAP likelihood
code.  We do not find the correlation function to be unusual beyond
the low quadrupole value. We develop a fast likelihood evaluation
routine that can be used instead of WMAP routines for low $\ell$
values.  We apply it to the Markov Chain Monte Carlo analysis to
compare the cosmological parameters between the two cases.  The new
analysis of WMAP either alone or jointly with SDSS and VSA reduces the
evidence for running to less than 1-$\sigma$, giving
$\alpha_s=-0.022\pm 0.033$ for the combined case.  The new analysis
prefers about 1-$\sigma$ lower value of $\Omega_m$, a consequence of
an increased ISW contribution required by the increase in the spectrum
at low $\ell$.  These results suggest that the details of foreground
removal and full likelihood analysis are important for the parameter
estimation from WMAP data.  They are robust in the sense that they do
not change significantly with frequency, mask or details of foreground
template marginalization.  The marginalization approach presented here
is the most conservative method to remove the foregrounds and should
be particularly useful in the analysis of polarization, where
foreground contamination may be much more severe.

  \end{abstract}


\maketitle

\section{Introduction}

Data analysis of cosmic microwave background maps is a challenging
numerical problem. The question that we want to answer is the
probability (or likelihood) of a theoretical model given the data.  In
order to evaluate the exact likelihood of a theoretical power spectrum
of CMB fluctuations given a sky map of these fluctuations it is
necessary to invert the theoretical covariance matrix. This operation
scales as $O(N^3)$, where $N$ is the length of the data vector and is
currently limited by practically available computer technology to $N
\lesssim 10^4$. One is hence forced to use approximate estimators when
inferring the power spectrum from data such as WMAP satellite
\citep{2003ApJS..148....1B}, which have 1-2 orders of magnitude more
independent measurements. The most popular methods are the pseudo-Cl
(PCL) method (see e.g. \citep{2002ApJ...567....2H}) and the Quadratic
Maximum Likelihood (QML) estimator (see e.g. \citep{1997PhRvD..55.5895T}). Both of these methods produce as an
intermediate step estimates of multipole moments $C_\ell$ and
approximate methods have been developed to describe their probability
distributions as accurately as possible
\citep{2000ApJ...533...19B,2003ApJS..148..195V}.  These perform
satisfactorily for high $\ell$ values, where the central limit theorem
guarantees a Gaussian distribution (in offset lognormal transformed
variables) will be a good approximation.  Unfortunately, these methods
are much less reliable at low multipoles, where the distributions are
not Gaussian. The situation is complicated further by the masks
applied to the data to remove the galactic foreground contamination
and by the marginalization of unwanted components, all of which makes
analytic approach unreliable.
In \cite{2003astro.ph..7515E} it was suggested to  use a hybrid approach using
QML on degraded maps at low $\ell$ and PCL at higher multipoles.

The issue of the exact values of multipole moments in WMAP data has
attracted much attention since the original analysis by WMAP team
\cite{2003ApJS..148..175S}. Several unusual features have been pointed
out already in the original analysis.  One of these was the
correlation function, which appears to almost vanish above
$60^{\circ}$.  Another was the low value of the quadrupole. With the
PCL analysis the value of the quadrupole was found to be $\sim 123 \mu 
{\rm K}^2$, compared to the expected value of $\sim 1200 \mu {\rm K}^2$ for the
simplest $\Lambda CDM$ model. The probability for this low value was
estimated to be below 1\%, depending on the parameter space of models.
The discussion of the statistical significance of the low values of
quadrupole and octopole in the WMAP data
\citep{2003MNRAS.346L..26E,2003JCAP...07..002C,2003JCAP...09..010C,PhLB..570..145L}
has sparked a renewed interest into the so called estimator induced
variance \citep{2004MNRAS.348..885E} - the error in the likelihood
evaluation arising due to the use of an estimator rather than the
exact expression. In \cite{2004MNRAS.348..885E} it has been argued that [QML
estimator performs significantly better than the PCL estimator and
that the true value of the quadrupole probably lies in the range
around $170-250 \mu {\rm K}^2$.  However, only the maximum likelihood value
was computed and not the full likelihood distributions so the
statistical significance of this result and its effect on the
parameter estimation remained unclear. In addition, the role of 
foregrounds and monopole/dipole removal has not been explored in detail. 

In this paper we take a different approach.  We argue that the actual
value of the best fitted quadrupole (and other multipoles) is not of
the main interest, since it can be quite sensitive to the details of
the foreground subtraction procedure, 
type of mask used and numerical details of the analysis
(in fact, the various values proposed so far may even be statistically
indistinguishable if the likelihood function at the peak is very
broad). What is more important is the probability or likelihood of a
model given the data, compared to another model that may, for example,
fit the data better.  This is encapsulated in the likelihood ratio
between models and within the Bayesian context is the only information
we really need to asses the viability of cosmological models that
belong to a certain class. In this paper we perform the exact
likelihood calculation by a direct inversion of the covariance matrix
for the low resolution maps, thus eliminating all the uncertainties
related to estimator variance approximations. Since we use low
resolution maps with less than 3000 pixels we can do the inversions
with a brute force linear algebra routines.  This means we cannot do
the analysis on all of the multipole moments, so we analyse low
multipoles with the exact method and use PCL analysis for the higher
multipoles, where the two methods agree with each other and where the
approximate variance estimates developed for PCL analysis are likely
to be valid.

Second issue we wish to address in this paper is the question of
foreground subtraction. This is done in two steps. First, pixels with
high degree of contamination are completely removed from the
data. This results in the so called KP2 (less aggressive, 85\% of the
sky) and KP0 (more aggressive, 75\% of the sky) masks
\citep{2003ApJS..148...97B}.  There remains contamination even outside
these masks in individual frequency channels. This contamination can
be further reduced using templates and/or frequency information
\citep{2003ApJS..148...97B}. In WMAP analysis the templates were
fitted for and subtracted out of WMAP data.  Even with a perfect
template there is a danger that this procedure can oversubtract the
foregrounds, since one is essentially subtracting out the maximum
amplitude consistent with the template which could include some of the
signal.  Instead, here we do not subtract out the templates, but
marginalise over them by not using any information in the data that
correlates with a given template.  
This procedure has not been applied to WMAP data in
previous analyses.
It 
guarantees that there is no statistical bias caused by the 
foreground removal. 

Some of the templates that were subtracted in WMAP analysis,
particularly 408MHz Haslam synchrotron radiation map
\citep{1982A&AS...47....1H}, are of poor quality. WMAP produced a
better set of templates applying Maximum Entropy Method (MEM) to WMAP
maps in several frequency channels using templates as priors only
\citep{2003ApJS..148...97B}. In addition to the Haslam synchrotron
map, they used \cite{2003ApJS..146..407F} H-$\alpha$ map as a tracer
of free-free emission and the SFD dust template based on
\cite{1998ApJ...500..525S}.  This process resulted in three MEM
derived foreground maps.  These, however, were not used to infer the
power spectrum. Instead, the official power spectrum was determined from the
integrated single frequency maps and the same templates that were used
as priors for the MEM map making procedure, ignoring the MEM derived
maps.

The MEM derived maps are likely to be the most faithful representation
of the foregrounds. When used in power spectrum inference, however,
they must be used with care due to complicated nature of their
signal and noise correlations
\citep{2003ApJS..148...97B}. Nevertheless, on the largest scales,
where receiver noise is negligible, they are probably the best
available option. We therefore use the integrated single channel maps
and the MEM derived foreground templates as a basis of our work. Note
that in foreground marginalization procedure no template is actually
removed from the data and there is no danger of introducing noise
correlations that could significantly affect the power spectrum
estimates. 
We perform this process on foreground unsubtracted maps of
the V and W channels of the WMAP satellite. We use both KP2 and KP0
masks and project out the remaining galactic contamination using MEM
inferred maps of dust, synchrotron and free-free foregrounds. We use
the likelihood evaluated in this way to asses the statistical
significance of departure from the concordant model at low multipoles
and to perform the statistical analysis of cosmological models given
the data.

WMAP team also produced the so called \emph{Internal Linear
Combination} (ILC) map of the CMB emission, by using internal maps at
various frequencies to decompose them into CMB and foreground
components.  This approach is not based on any templates and so uses
less information than in principle available. While visually these
maps appear to be relatively free of contamination outside the
galactic plane, there are still artifacts within the plane. This means
that one must be careful when projecting out monopole and dipole: one
should not simply remove them from the all-sky map, since they could
be contaminated by galactic emission at the canter and this would
leave a residual offset outside the galactic plane, which could
contaminate all of low multipoles. One must again apply the
marginalisation over monopole and quadrupole on the masked map to
eliminate any contamination in the final result. A similar
approach has been taken by \cite{2003PhRvD..68l3523T}  and \cite{2004astro.ph..3098E}, who produced
their own versions of ILC maps. Since we argue that the best method is
to use single frequency maps together with correct templates and we use
ILC map for illustration and cross-check purposes only, we do not
consider these alternative ILC solutions further.

\section{Method}

\subsection{Likelihood evaluation}

Given noise-less and independent measurements of the CMB sky $\mathbf{d}$, the theoretical
covariance matrix for these measurements is given by \citep{1998PhRvD..57.2117B}
\begin{equation}
  C_{i,j} = \sum_{\ell=2}^{\infty} \frac{2\ell+1}{4\pi}  C_\ell
  P_\ell(\cos \theta_{i,j}), 
\label{eq:1}
\end{equation}
where $C_\ell$ is the power spectrum, $P_\ell$ is the Legendre Polynomial of order
$\ell$ and $\theta_{i,j}$ is the angle between $i$th and $j$th point
on the sky. We also define 
\begin{equation}
  \CC_\ell = \frac{C_\ell \ell (\ell+1)}{2\pi},
\end{equation}
which is the quantity that is conventionally plotted (and often
referred to) as the power spectrum. 

In addition to the covariance matrix in equation \ref{eq:1} we want to
project out linear components of the data vector that correspond to
known contaminants in our data. Fortunately, there exist a standard
procedure for this \citep{1992ApJ...398..169R}: the covariance matrix
of the contaminant is calculated and added to the theoretical
covariance matrix with a very large variance. 
Here the covariance matrix of the template is given by 
$\bi{C=\langle LL^{\dagger}\rangle}$, where $\bi{L}$ is the 
template vector. 
Using this method, we
project out the map's monopole, dipole and the known galactic
contaminants, namely dust, synchrotron and the free-free emission. For
completeness we add the diagonal noise component $N_{ii}=\sigma_i^2$,
although this is not strictly required for this analysis, because the
noise power spectrum is $<10\mu {\rm K}^2$ on scales of our interest.

Hence, the  total covariance matrix can be written as
\begin{eqnarray}
  \mathbf{C}^{\rm total}  = \mathbf{C} + \mathbf{N} +  
\lambda ( \mathbf{C}^{\rm
  dust}+\mathbf{C}^{\rm synch}+\mathbf{C}^{\rm
  free-free}&+& \nonumber \\ \mathbf{C}^{\ell=0}+\mathbf{C}^{\ell=1} )&.&
\end{eqnarray}
The value of $\lambda$ in the above equation must be large enough so
that unwanted components are projected out. If it is too large,
however, the numerical errors start to affect the results.

The logarithm of likelihood of given $C_\ell$s  can then be written as
\begin{equation}
  \log L = -\frac{1}{2} \mathbf{d^T (C^{\rm total})^{-1} d} - \frac{1}{2}\left(\log
  |\mathbf{C}| + N \log 2\pi\right),
\label{eq:2}
\end{equation}
where $\bi{d}$ is the data vector. 
To evaluate the likelihood of a given theoretical model we simply evaluate this expression,
computing the covariance matrix using the theoretical model 
spectrum $C_\ell$ in equation \ref{eq:1}. 

\subsection{Choice and preparation of maps}


As mentioned in the introduction the procedure described above can
realistically be performed only on modestly sized maps. We decrease
the resolution of a given map using the following procedure: Firstly,
the full resolution source map is multiplied by the mask, whereby
every masked pixel is zeroed, while unmasked remaines the same. The
map is then smoothed by the $5^\circ$ FWHM Gaussian beam and resampled
at a lower Healpix \citep{2002adass..11..107G} resolution (\texttt{nside}=16), giving 3072 roughly
independent pixels on a full-sky map. The mask itself is smoothed in
the same manner and this gives us information by how much the smoothed
pixels that were affected by the mask need to be up-scaled. We do not
use pixels whose smoothed mask value drops below 0.7. We use this
information to reduce the effective scale of smoothing beam (by square
root of this correction) in the calculation of covariance matrix,
although we verified that this does not affect any of the final
results.

We have also attempted an exact calculation of the window function
treating each subpixel of a low resolution Healpix map
separately. Unfortunately, this is computationally prohibitively
expensive. Instead, we have performed weighted averaging within each
low-resolution Healpix pixel using the effective Healpix window
function provided with the package and get compatible results. 
We chose not to adopt this approach for the main analysis since the 
individual Healpix pixel windows are anisotropic, depend on the mask and are very 
slowly dropping off with $\ell$. For our resolution level the effective 
windows (which are only valid for full sky coverage) 
are only given up to $\ell=64$ and there is still a lot of power beyond that. 

The gaussian smoothing
procedure is used on WMAP integrated maps for V and W channels
and for the MEM maps for the three major foregrounds: Dust,
Synchrotron and Free-Free emission. In all cases, the low resolution
maps were produced for the KP2 mask and for the more conservative KP0
mask. We also applied the same procedure to the ILC map, except that 
in this case we smooth over the whole map and so do not need to 
upscale the pixel values by the effect of the mask. 
By changing various parameters of the inversion process we get consistent
likelihoods and we estimate the uncertainty in likelihood evaluation to
be about 0.2 in logarithm of the likelihood. 

\section{Multipole moments and their statistical significance}

\begin{figure}
\epsfig{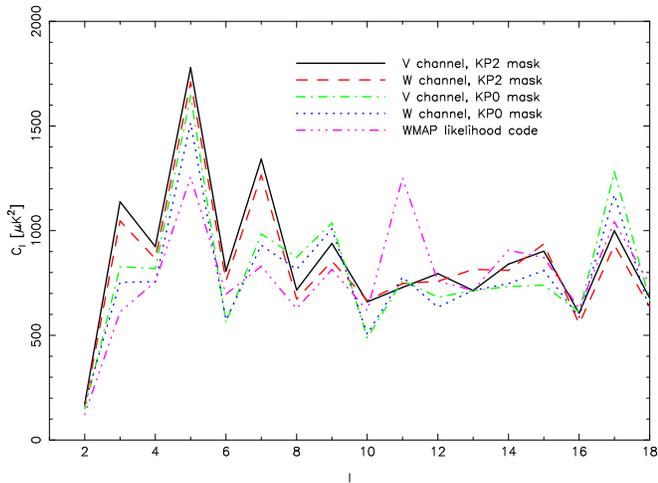}

\caption{
This figure shows the maximum likelihood power spectrum for
  several combinations of frequencies and masks. Note that all spectra
  agree reasonably well beyond $\ell=11$.
\label{fig:cls} }
\end{figure}

In figure \ref{fig:cls} we show the maximum likelihood (ML) values of the
multipoles up to $\ell=18$ for several of our basic cases. One can see from
this figure that most of the estimates up to $\ell \sim 10$ are above the
PCL values given by WMAP team, while at higher multipoles the two
agree well ($\ell=11$ appears anomalous and PCL gives 
a much higher value than the exact likelihood
analysis). Some of the differences are due to random fluctuations:
KP2 mask contains 85\% of the sky compared to 75\% for KP0 and this
can lead to differences in the two estimates.  Similarly, projecting
out the foreground templates reduces the amount of information, so
there can be statistical differences between our analysis and the one
without marginalization.  
While the differences between KP0 and KP2
masks are likely to be within the allowed range of statistical
fluctuations, this is less likely for the differences between WMAP-PCL
and our analysis of V with KP2 mask, since the same mask and channel 
have been used by WMAP.  The difference is partly due
to the use of the exact likelihood analysis and partly due to the
foreground marginalization.
While WMAP team marginalized over monopole and dipole, 
they subtract out
the foregrounds with the maximal amplitude, which may have
removed some of the true cosmological signal and pushed the values lower.  
To eliminate the bias
that can arise from this procedure it is best to exclude the
information in the signal that correlates with templates and with 
monopole/dipole. This reduces
the statistical power, but is guaranteed to be unbiased.

\begin{figure}
\epsfig{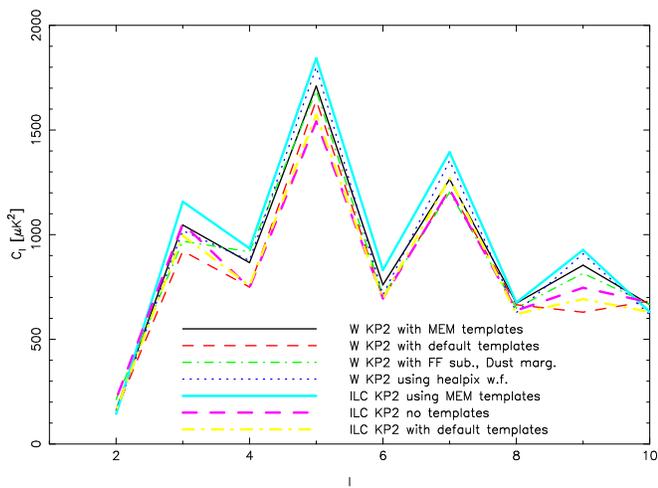}

\caption{This figure shows the maximum likelihood power spectrum up to
  $\ell=10$ for several test cases. The derived features in the most
  likely values of power spectrum are robust with respect to choice of
  window function (gaussian versus healpix), templates (MEM versus 
external, dust only versus standard 3 templates in W) and maps (W versus ILC). 
\label{fig:cls2} }
\end{figure}

To investigate further the robustness of our results 
figure \ref{fig:cls2} shows the most likely values of
power spectra (up to multipole $\ell=10$) for various combinations of
template choice for W channel data and for the ILC map. This
realistically indicates potential systematic differences arising due
to choice of templates. On the same graphs we also show the reduction
performed with Healpix window functions (where pixels of the
low-resolution map were only weighted averages of all corresponding
pixels in the high-resolution map), indicating that our results are
robust with respect to choice of window function and smoothing
procedure. The differences between the various cases are small 
compared to the difference between exact evaluations and WMAP 
values. 
We emphasize that ML values are the least robust 
part of the analysis and it is much more important that the 
probability distributions away from the peak are consistent. For 
quadrupole we discuss this below, while the overall impact on 
the cosmological parameter estimation 
is discussed in next section. 

Marginalisation procedure is guaranteed to give unbiased results 
independent of the form of the template or its 
nongaussian properties. The only assumption is that the template
is not correlated with the true CMB, which could happen if the 
templates are produced from the CMB data itself and were affected 
by noise, calibration or beam uncertainties. It is unlikely that 
this would happen on large scales. We have tested this 
hypothesis by using the extrenal templates instead of MEM
templates, without finding much differences in the result (figure \ref{fig:cls2}). 

\begin{figure}
\epsfig{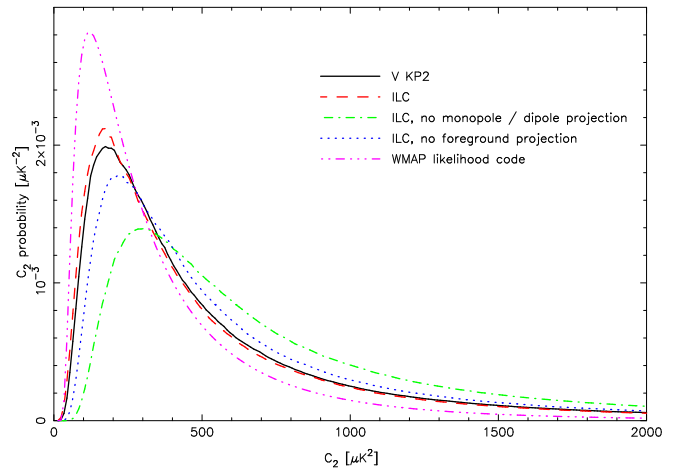}

\caption{
This figure shows the probability distributions for the value
  of $\CC_2$ as inferred for the various combinations of the 
monopole/dipole and the
  foregrounds marginalization.
  Also shown is the official WMAP likelihood code 
output.
Note that $\CC_2$ in V without monopole/dipole or the
  foregrounds marginalization is heavily contaminated and 
gives very high ML values, while all the other cases 
have very similar probability distributions
(except WMAP code).
\label{fig:marg} }
\end{figure}

Figure \ref{fig:marg} shows the probability distribution for
the value of $\CC_2 $ for several cases, assuming best fit $\Lambda
CDM$ model for other $\CC_\ell$s. Particular choice of other $\CC_\ell$s
affects the inferred curves, although at a level below the variance
between various curves. 
It has several interesting
features. Firstly, when all marginalizations are used, the V channel,
W (not shown)
and ILC give very consistent results. In the absence of 
marginalization and foreground subtraction
the V and W channel maps are
very affected by foregrounds and ML values reach up to 500$\mu {\rm K}^2$. 
The ILC map could  
be affected by the foreground marginalization; its value
drops from $\sim 220 \mu {\rm K}^2$ (consistent with
\cite{2004MNRAS.348..885E}) to $\sim 170 \mu {\rm K}^2$ when projection is
included in analysis. The ILC map may suffer even more from the residual
monopole / dipole contamination, which  pushes the quadrupole value up.  

\begin{figure}
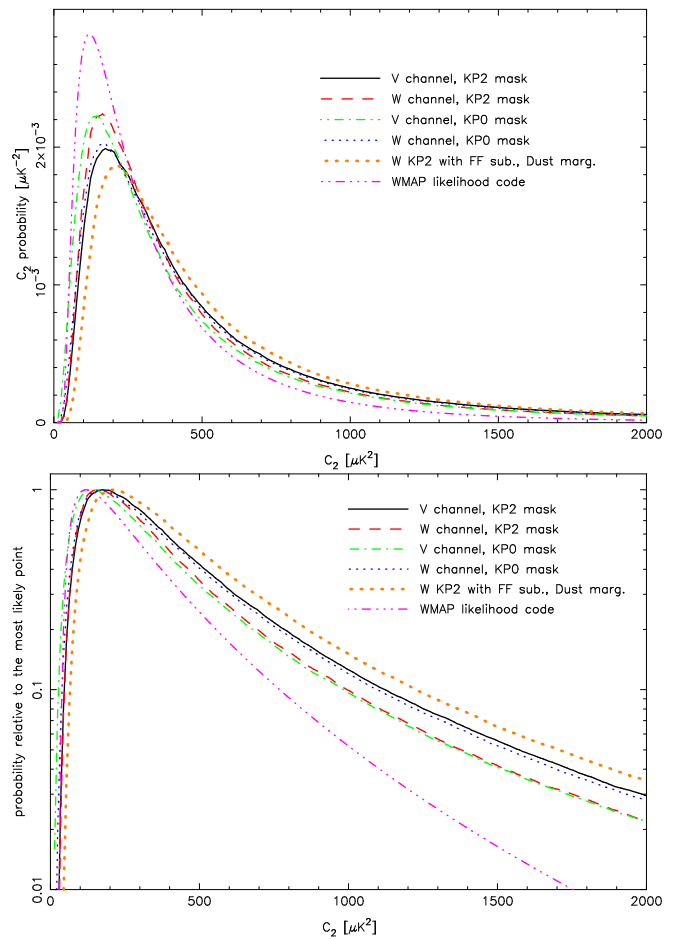

\epsfig{file=realtom0.ps, height=\linewidth, angle=-90}

\epsfig{file=realtom.ps,  height=\linewidth, angle=-90}

\caption{This figure shows the probability distributions for the value
  of $\CC_2$ as inferred for the various combinations of the selected
  channel and mask and the official WMAP likelihood code. The upper
  panel shows the normalised probability distribution, while the lower
  panel shows the probability relative to the most likely point. Note
  that the lower panel's vertical axis is logarithmic. Values of other
  $C_\ell$s were set to those of the best fit $\Lambda CDM$
  model. \label{fig:realtom} }
    
\end{figure}

While our procedure of marginalising over 3 templates is the most 
conservative, one may worry that it is unnecessary. Some of the 
channels are not really strongly contaminated by all 3 components and
if frequency scaling is known then 
multi-frequency information can be used to constrain a given component
in a given channel. 
While there is nothing wrong with our procedure one could argue that it reduces
the amount of information. The number of eliminated modes is roughly given 
by the number of templates used, but since the templates are correlated
(being all dominated by our galaxy) the information loss
from  large scale modes is likely to be less than 3.
It is also not clear how the templates couple to different multipoles. 
To test these effects we perform the analysis in W channel, where 
foreground contamination is dominated by dust. We 
use marginalization only over SFD dust template (subtracting out the 
free-free component and doing nothing for synchrotron).  
We find this has very little effect on the maximum likelihood values of 
multipoles, as
shown in figure \ref{fig:cls2}. 
For $\ell=2$ we find ML value at $220 \mu {\rm K}^2$, slightly 
higher than in other cases (figure \ref{fig:realtom}), but the 
overall probability distribution is very similar 
to other cases. The effect of this procedure on the
parameter estimation is explored in the next section. 

It is interesting to asses the statistical significance of the
departure of the lowest multipoles from the concordant model. Our
focus is not on the actual statistical procedure of assessing this
departure (see e.g. \citep{2003MNRAS.346L..26E}), but on the effect
of estimator induced variance. We consider 5 cases: all possible
combinations of the choice of mask (KP2 or KP0) and frequency (V
channel or W channel) and the official WMAP likelihood code
\citep{2003ApJS..148..195V,2003ApJS..148..135H}.
The inferred maximum likelihood values (figure \ref{fig:realtom}) lie in
the range $140\mu {\rm K}^2- 220 \mu {\rm K}^2$, but the likelihood function 
is broad at the peak and the exact value of the maximum likelihood estimate 
is driven by small details in the analysis: in all of our basic cases the 
likelihood is within 10-20\% of the peak value 
over the range (120-250)$\mu {\rm K}^2$. Thus 
our results are consistent both with the original 
WMAP value ($123 \mu {\rm K}^2$) and the values in \cite{2004MNRAS.348..885E} and 
there is no ``correct'' value given the level of foreground contamination.

As we argued in introduction, the precise value of maximum likelihood
estimator is not of primary interest, given that it can be strongly
affected by the details of the analysis. Much more important for the
question of parameter estimation is the shape of the likelihood
function. Figures \ref{fig:marg}-\ref{fig:realtom} shows that while the maximum
likelihood value of the quadrupole is quite uncertain, 
all of our
cases give very similar shapes of the likelihood function. 
This likelihood distribution is not consistent with the
likelihood provided by WMAP team,  which appears to underestimate the
errors associated with the galactic cut and marginalizations. The WMAP
likelihood of the concordant model $\CC_2$ ($\sim 1200 \mu {\rm K}^2$) is
roughly 2.5 times too low with respect to the most likely point when
compared to our likelihood values. 
This change
in the shape of the likelihood function affects the parameter
estimation, particularly the running of the spectral index, 
as shown in section {\ref{sec:params}}.  
We note here that not performing the marginalization over foregrounds and/or 
monopole/dipole would lead to an even 
higher probability of concordance model compared to low $\CC_2$ models
(see the corresponding probability distributions in figure \ref{fig:marg}), 
but these are more likely to be contaminated and should not be used in the 
likelihood analysis.   

\begin{figure}
\epsfig{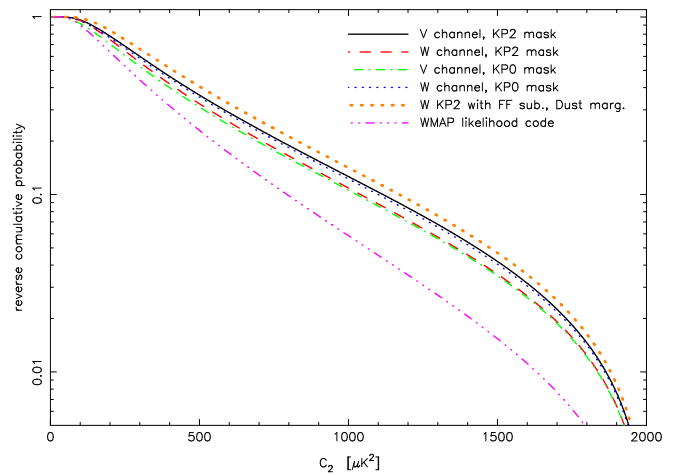}

\caption{ Cumulative probability as a function of the true value of
the quadrupole (integrated from large values downwards assuming $0 <
\CC_2 < 2000 \mu{\rm K^2}$).  \label{fig:p1} }
    
\end{figure}

Figure \ref{fig:p1}
shows the integrated probability as a function of the true value of
the quadrupole (integrated from large values downward), under the
assumption that the prior distribution of quadrupole values is
uniform between 0 and 2000$\mu {\rm K}^2$. This
prior is is adopted due to the fact that the concordance value of
quadrupole is $\sim 1000\mu {\rm K}^2$ (see e.g.  \citep{2003MNRAS.346L..26E}). This gives the probability of the true value exceeding
$\CC_2$ assuming this prior.  We find that this probability is around
10\%, as opposed to 4\% by the WMAP likelihood analysis. Thus with
uniform prior on values of $\CC_2$ the probability of the true
quadrupole to be above that predicted by the concordance model
is not particularly small. It becomes even larger if the upper limit 
at 2000$\mu {\rm K}^2$ is removed, in which case we find 18\% probability of 
the true value exceeding the concordance value.

Note that WMAP team chose to give the statistical significance of the
low quadrupole in terms of number of random realizations of
theoretical models in Monte Carlo Markov Chains (MCMC) for which the
extracted quadrupole is lower than the observed value of $123\mu {\rm K}^2$.
This is a frequentist statistic which cannot be directly
compared to the one we defined here in the context of Bayesian
statistics.  The frequentist approach leads to lower numbers (less
than 1\%, compared to 4\% above) for the specific value of the
quadrupole obtained by WMAP, but the probability is likely to be
higher if our analysis procedure was applied to the data given our
broader likelihoods and higher values of the best fitted quadrupole.
The WMAP analysis does not include the uncertainties in the 
foreground subtractions, which should have an important effect given the 
skewed nature of the probability distributions: if 
an error estimate of 50$\mu {\rm K}^2$ on $\CC_2$ were added to the 
measured value it 
would lead to an increased probability of concordance model. 
In order to truly decouple the cosmic variance uncertainty for the
errors arising from the galactic cut and foregrounds one would need to
infer the probability distribution for a particular realisation of
$a_{\ell m}'s$. For a full-sky CMB observation with no galactic
contamination, this would be a delta function; galactic cut and large
scale contamination would spread the
probability over a finite region. This distribution, marginalised to
produce $p(\langle a_{2m}^2 \rangle_{m})$ would be the correct quantity that must be
compared to the concordant value and the corresponding cosmic
variance. Work on this front is currently in progress.

While the frequentist approach does allow one to test a model (or a
class of models) independent of other models, it is still not free of
assumptions.  Testing the quadrupole on its own only makes sense if we
believe that there is something special about it, for example because
it is sensitive to the physics on the largest scales, which may not be
probed by lower multipoles.  If it is not viewed as special, but only
one of the many estimated multipoles, then the probability of one of
them being this low is significantly higher.  This is tested in the
frequentist approach with the goodness of fit ($\chi^2$), which for
WMAP does not reveal any particular anomalies.  Unfortunately there is
no hope to resolve these statistical questions completely with only
one observed sky.

\begin{figure}
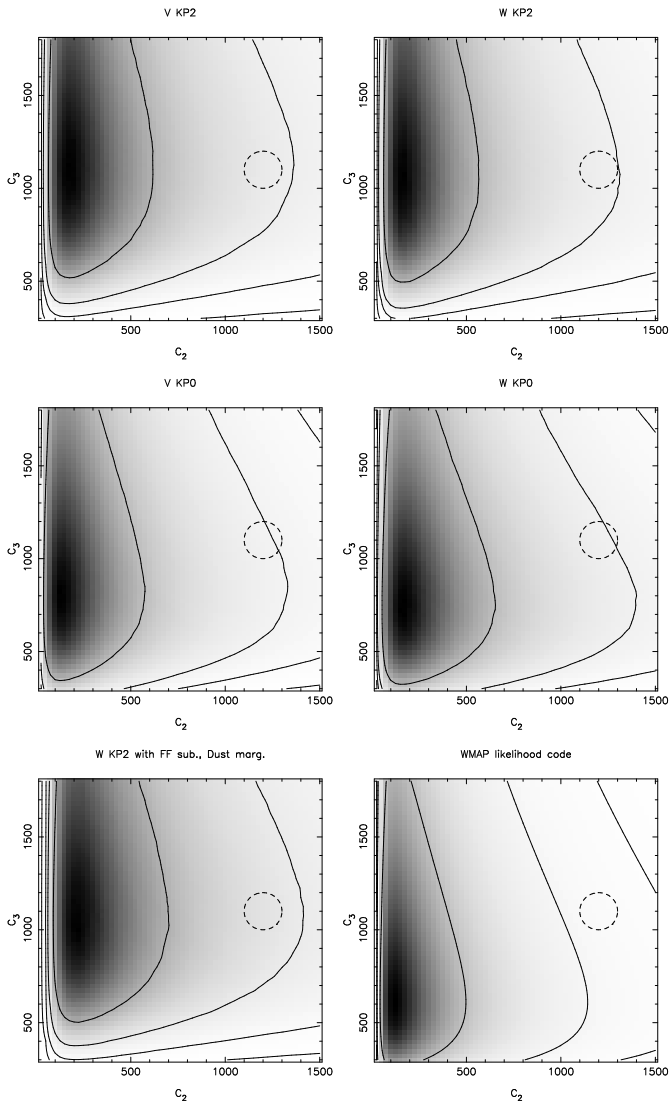

\begin{tabular}{cc}
\epsfig{file=g1.ps, height=0.5\linewidth, angle=-90}&
\epsfig{file=g2.ps, height=0.5\linewidth, angle=-90}\\
\epsfig{file=g3.ps, height=0.5\linewidth, angle=-90} &
\epsfig{file=g4.ps, height=0.5\linewidth, angle=-90}\\
\epsfig{file=gx.ps, height=0.5\linewidth, angle=-90}&
\epsfig{file=g5.ps, height=0.5\linewidth, angle=-90}
\\
\end{tabular}

\caption{\label{fig:c2c3}. In this figure we show the probability
  distribution function on the $\CC_2$-$\CC_3$ plane for all
  considered possibilities and the original WMAP likelihood
  code. Contours correspond to the one, two, three and four sigma
  assuming top-hat priors on the plotted limits ($0<\CC_2<1500
  \mu {\rm K}^2$,$300<\CC_2<1800 \mu {\rm K}^2$). The dashed circles correspond to
  the approximate values of the concordant model. }

\end{figure}

In figure \ref{fig:c2c3} we plot the contour plots of parameters on
the $\CC_2$-$\CC_3$ plane for the considered models. This shows that
the likelihoods between $\CC_2$ and $\CC_3$ are only weakly
correlated, both for exact likelihood evaluation as well as for the
PCL approximation. In original analysis there was some evidence for both 
$\CC_2$ and $\CC_3$ being low, so that the overall significance was 
between $2-3 \sigma$  (figure \ref{fig:c2c3}). 
The evidence for discrepancy weakens
below $2 \sigma$ with our analysis and is consistent among the four cases.

\begin{figure}
\epsfig{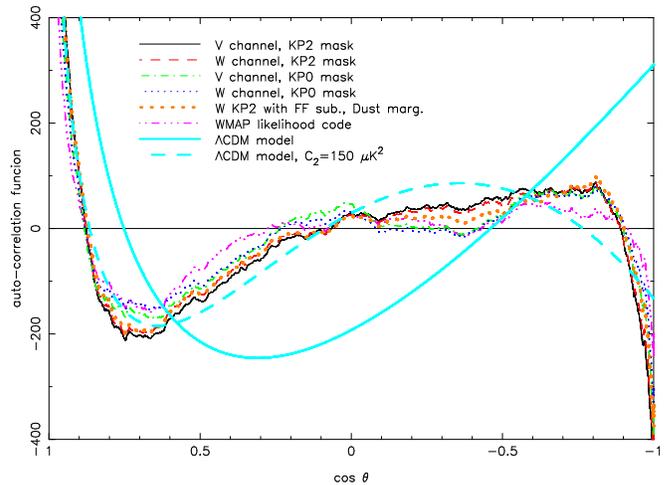}

\caption{This figure shows the autocorrelation function for all
  considered cases, the $\Lambda CDM$ model favoured by the WMAP data
  and the same model with $\CC_2$ set to $150 \mu {\rm K}^2$.\label{fig:autocor}.  }

\end{figure}

WMAP team presented further evidence of unusual nature of large scale
correlations using the correlation function, which appears to vanish
on angles above $60^{\circ}$ \citep{2003ApJS..148..175S}.  Correlation
functions are notoriously difficult to interpret due to the correlated
nature of the values at different angles, so one must be careful not
to over-interpret such results.  In figure \ref{fig:autocor} we show
the correlation function analysis for these cases, compared to the
original WMAP analysis and to theoretical predictions of $\Lambda CDM$
model. We also show the result for the $\Lambda CDM$ model where
$\CC_2$ has been lowered to $150 \mu {\rm K}^2$, keeping the other
multipoles unchanged.  Several features are apparent from this
figure. First, theoretical predictions for large scale correlation
function are largely driven by the quadrupole and lowering its value
to $150 \mu {\rm K}^2$ brings the correlation function into a significantly
better agreement with the observations than the unmodified $\Lambda
CDM$ model. Second, our results significantly modify the predicted
correlation function and the deviations from zero on large angles are
now much more evident, both in the positive direction and in the
negative direction at very large angles.  To investigate it further
WMAP team introduced a statistic $S=\int_{-1}^{0.5}[C(\theta)]^2d\cos
\theta$.  This is a posteriori statistic that was designed to maximise
the effect, so its statistical significance is difficult to
evaluate. We find that its value increases from 1691 for WMAP analysis
to 4197 (W KP0), 5423 (V KP0), 9086 (V KP2), 7698 (W KP2) and 5832 (W KP2, 
dust marginalization only).  While its value for
standard $\Lambda CDM$ model is 49625, reducing the quadrupole to $150
\mu {\rm K}^2$ changes this to 8178, below the value we find in the case of
V KP2.  We conclude that there is no obvious anomaly in the
correlation function beyond the fact that the quadrupole is low and
there is no evidence of the correlation function vanishing on large
angles.

\section{Parameter estimation}
\label{sec:params}

In order to asses the effect of the exact likelihood evaluations on
the inferred cosmological parameters we have run the Markov Chain
Monte Carlo parameter estimations using the original WMAP likelihood
code and the code modified to use the exact calculations at lowest
multipoles. The total likelihood was calculated by evaluating the
likelihood for the $\ell \le 12$ multipoles using the exact matrix
inversion and adding the likelihood evaluated from the remaining
multipoles using the WMAP likelihood code.  The power spectrum values
for the multipoles $\ell>12$ in the exact likelihood code were kept at
the WMAP PCL most likely model when calculating the covariance matrix:
this ensures that the likelihood is not ``accounted for'' twice. We
also neglect the anti-correlation between $\ell=12$ and $\ell=13$
modes at the boundary. The evaluation of the exact likelihood
typically takes around a few seconds on a modern workstation and this
is less than the time it takes to evaluate a theoretical CMB power
spectrum with CMBFAST \citep{1996ApJ...469..437S}.  Therefore, using
the exact likelihood code does not slow down the MCMC parameter
estimation significantly. Each of the chains described below contains
100,000-200,000 chain elements, the success rate was of order 30-60\%,
correlation length 
10-30
and the effective chain length of order 5,000-15,000.
We use 8-24 chains and
in terms of Gelman and Rubin $\hat{R}$-statistics \citep{gelman92} we find the 
chains are sufficiently converged and mixed, with $\hat{R}<1.01$,
compared to recommended value $\hat{R}<1.2$ or more 
conservative value $\hat{R}<1.1$
adopted by WMAP team \citep{2003ApJS..148..195V}. 

The likelihood also uses the information contained in the
  polarization-temperature (TE) cross-correlation power spectrum using
  the official WMAP likelihood code, which uses similar approximations
  as temperature power spectrum and completely ignores correlations
  between TT and TE power spectra. We cannot yet use the
  exact evaluations since the polarization maps are not publicly
  available at this time.

We ran several MCMCs using a custom developed software described in
\cite{2003MNRAS.342L..79S}.  We consider only flat models.
We begin with the simplest
5-parameter models
\begin{equation}
\bi{p}=(\tau,\omega_b,\omega_{\rm cdm},{\cal R},\Omega_m), 
\label{5p}
\end{equation}
where $\tau$ is the optical depth, $\omega_b=\Omega_bh^2$ is
proportional to the baryon to photon density ratio, $\omega_{\rm
cdm}=\Omega_{\rm cdm}h^2$ is proportional to the cold dark matter to
photon density ratio, $\Omega_m=\Omega_{\rm
cdm}+\Omega_b=1-\Omega_{\lambda}$ is the matter density today and
${\cal R}$ is the amplitude of curvature perturbations at $k=0.05$/Mpc
(we replace this parameter with $\sigma_8$ in table
1). 
To reduce the degeneracies we use 
$\omega_b$, $\omega_{\rm cdm}$, angular diameter distance $\Theta_s$, 
$\ln {\cal R}$ and $\ln {\cal R}-\tau - 0.5\log (\omega_b+\omega_{\rm cdm})$ 
instead of parameters in equation \ref{5p}, adopting broad flat priors on them. 
Most of
these priors are not important because the parameters are well 
determined. The exception is optical depth, for which we 
additionaly apply $\tau<0.3$
on some of MCMCs following WMAP team.  

The simple 5-parameter model is sufficient to obtain a
good fit to the WMAP data.  We add CBI+ACBAR to the WMAP data
\citep{2002astro.ph..5384M,2002astro.ph.12289K} and follow WMAP team
in denoting this dataset as WMAPext.  Second set of MCMCs we ran was
also based on WMAPext data, but with an expanded set of parameters
which include primordial slope $n_s$, its running $\alpha_s=dn_s/d\ln
k$ and tensors (parametrised with $r=T/S$), adopting flat priors on 
these parameters.  Adding these 3 parameters
only improves $\chi^2$ by 5, so they are not really needed to improve
the fit to the data.  Because of this we find significant degeneracies
among many of the parameters. The best fitted values are not
necessarily very meaningful and they could be significantly influenced
by the assumed priors, but we can still compare the changes between
the new and original analysis.  Third set of MCMCs was based on the
combined WMAPext+SDSS analysis \citep{2003astro.ph.10725T}, which
breaks some of these degeneracies. Last set of MCMCs was based on 
WMAP+VSA \cite{2004astro.ph..2498D}, both with and without SDSS. 
We remove $\tau<0.3$ constraint for this case. 
The results are shown in tables 1-2. 

\subsection{Matter density}
\begin{figure}
\epsfig{file=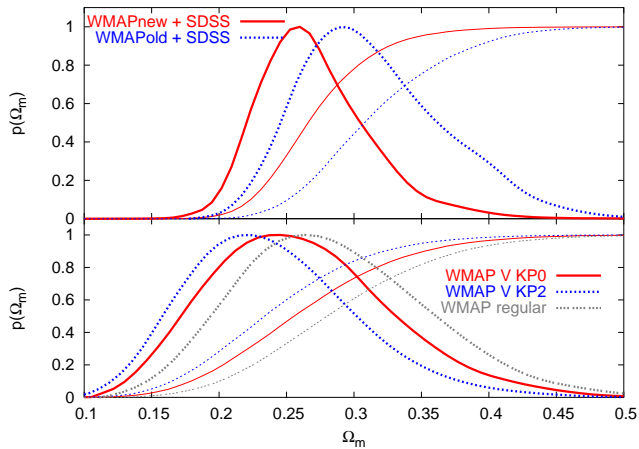, width=\linewidth}

\caption{
Probability distribution $p(\Omega_m)$ and its cumulative value 
$\int_{-\infty}^{\Omega_m} p(\Omega_m')d\Omega_m'$ for 
5-parameter MCMCs of WMAPext data (bottom) and for 8-parameter MCMCs
of WMAPext+SDSS data (top).
We present V frequency map and both KP0 and 
KP2 mask results for the full likelihood analysis of 5-parameters MCMCs 
of WMAPext data
and V KP2 for full likelihood analysis of 
8-parameter MCMCs of WMAPext+SDSS data. Also shown for comparison are
the results using regular (old) WMAP analysis routine.
\label{fig:om}
}
    
\end{figure}

In 5-parameter chains
$\Omega_m$ is the parameter that 
changes most by the new analysis.  Its probability
distribution from various MCMCs is shown in figure \ref{fig:om}. This
parameter is not well determined from the CMB data, since it only
weakly affects the positions of acoustic peaks in a flat
universe. This leaves the integrated Sachs-Wolfe effect on large
scales as an important way to constrain $\Omega_m$: reducing
$\Omega_m$ leads to a decay in the gravitational potential, which
increases the contribution to the large scale anisotropies from the
line of sight integration of the time derivative of gravitational
potential. Increase of the low multipoles by our analysis (figure
\ref{fig:om}) thus requires a lower value of $\Omega_m$ to fit the
data. This is more prominent for KP2, where the best fit value is 
$\Omega_m=0.24^{+0.07}_{-0.05}$, than KP0 which gives $\Omega_m=0.26^{+0.07}_{-0.06}$, 
but the latter contains
less area and its error distribution is slightly broader. Lower $\Omega_m$
values are also preferred in the joint WMAPext+SDSS analysis, but here
the SDSS data tend to push the overall value up to 
$\Omega_m=0.27^{+0.05}_{-0.03}$. In these 8 parameter
chains the WMAP $\chi^2$ is higher by about 5 compared to the WMAP
without SDSS. Thus there is a bit of a tension between the SDSS data,
favouring high $\Omega_m$ and the WMAP data favouring low values of this
parameter, although the statistical significance of this tension is
low.
For low $\Omega_m=0.24$ the 
Hubble parameter is $h=0.75$, still in agreement with the HST key project 
value of $h=0.72\pm 0.08$ \citep{2001ApJ...553...47F}.
If we eliminate tensors from the analysis then we find 
$\Omega_m=0.30^{+0.06}_{-0.05}$
for WMAP+SDSS+VSA combination of the data. The overall conclusion is that values of 
$\Omega_m$ between 0.2 to 0.4 remain acceptable by the data and that 
the actual value depends strongly on the choice of parameter space.

\begin{figure}
\epsfig{file=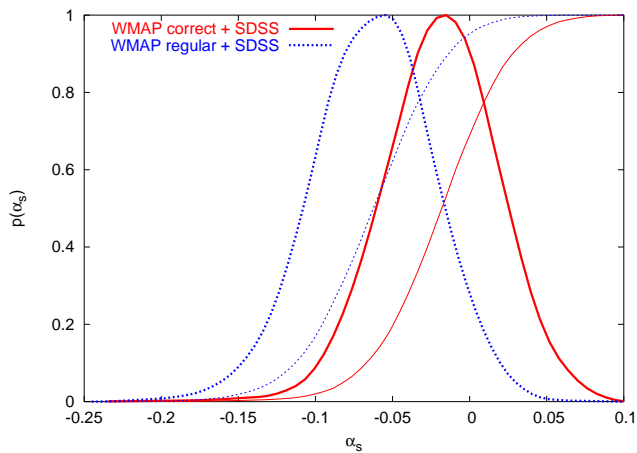, width=\linewidth}

\caption{
Probability distribution $p(\alpha_s)$ and its cumulative value 
$\int_{-\infty}^{\alpha_s} p(\alpha_s')d\alpha_s'$ for old 
and new MCMCs using WMAPext+SDSS data.
We use V frequency map and KP2 mask in the full likelihood analysis. 
\label{fig:running}
}
    
\end{figure}

\subsection{Running}

\begin{figure}
\epsfig{file=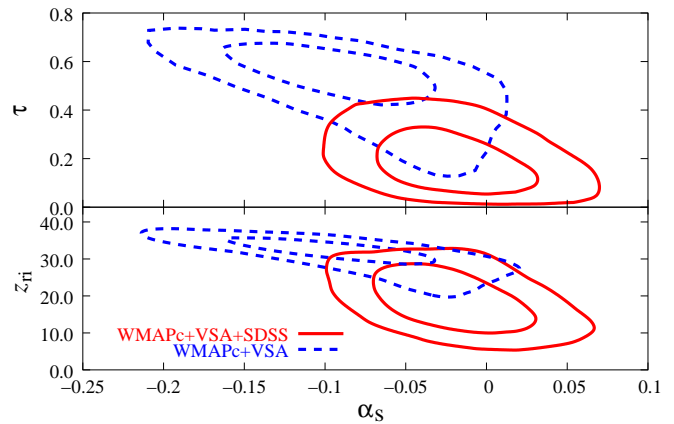, width=\linewidth}

\caption{Two dimensional contours of 68\% and 95\% probability 
in $(\alpha_s,\tau)$ and $(\alpha_s,\rm{z_{ri}})$ plane 
from WMAP+VSA and WMAP+VSA+SDSS data.
\label{fig:alphastau}
}
    
\end{figure}

Running has attracted a lot of attention ever since WMAP team argued
for a 2-$\sigma$ evidence of negative running.  When analyzing CMB
data alone one finds that running is strongly correlated with the
optical depth $\tau$.  Figure \ref{fig:alphastau} shows an example of
this in WMAP+VSA MCMCs. We see that this particular combination of
data prefers $\tau>0.3$ and that such a high value of optical depth
requires large negative running.  A similar effect has been noticed in
WMAP+CBI analysis \citep{2004astro.ph..2359R} and WMAP+VSA
  analysis \citep{2004astro.ph..2466R}.  We find that the
statistical significance of running is strongly affected by the
adopted prior on $\tau$. 
In fact, when prior on $\tau$ is relaxed, the one-dimensional
marginalised probability distribution seem to prefer models with high
values of $\tau$ and large negative running. However, we note that this is the result
of the large posterior probability volume in this region,
rather than a better fit to the data. Moreover, such high
values of optical depth are difficult to reconcile with the
hierarchical models of structure formation and would require a lot of
small scale power, contrary to the effect of a negative running.  Even
more importantly, a high optical depth would lead to a large signal in
WMAP $EE$ polarization spectrum. To eliminate this region of parameter
space WMAP team adopted a prior $\tau<0.3$ and we follow that for most
of our MCMCs.  However, one can also eliminate this region of
parameter space by adding the SDSS data, which do not favor the high
optical depth values (figure \ref{fig:alphastau}) and we give an
example of this in table 2.

In this paper we are more interested in how running changes
if we use the exact likelihood routine as opposed to the approximate one.
The resulting values of the running for various cases 
are given in tables 1-2. They
are significantly affected by the exact likelihood 
calculations. This is expected from the analysis presented in 
previous section, where we have shown that the 
exact likelihood analysis with foreground marginalization leads 
to an enhancement of low $\ell$ multipoles and broadens the  
shape of the likelihood 
distribution for quadrupole to allow a higher likelihood for models with less
negative running. 
Figure \ref{fig:running} shows the MCMC generated probability
distributions for running $\alpha_s$ using
WMAPext+SDSS in 8-parameter models. Note  
that there is a strong
correlation between running and tensors in such a way that for no
tensors there is less evidence for running
\citep{2003MNRAS.342L..79S}.  So some of the evidence for running in
the 8-parameter analysis (and in
\cite{2003ApJS..148..213P}) is driven simply by the large parameter
space of $r>0$ models and should not be taken as an evidence of running on its own.  
Even so we find that the evidence for
running, marginally suggested by the old analysis, largely goes away
in the new analysis and the value of running 
changes from -0.060 to -0.015 (V KP2, full marginalization) or -0.032 
(W KP2, dust marginalization only), with an error of 0.035. 
This confirms that the suggested evidence for
running relies crucially on low quadrupole and octupole
\citep{2003MNRAS.342L..72B}, for which the statistical analysis and
foreground removal are least reliable.  

This point was also noted in the recent analysis of WMAP+VSA data \citep{2004astro.ph..2466R}, where
the WMAP likelihood code was used and evidence in excess of 2-$\sigma$ for 
running was found, while removing $\ell<10$ information reduced this 
evidence to less than 1-$\sigma$. While one should not simply remove 
the entire $\ell<10$ information  one should use the exact calculations instead 
of approximate ones if the answer depends on it. 
Our results for WMAP+SDSS+VSA analysis for 7-parameter models without 
tensors given in table 2 show that running 
is strongly suppressed with the new analysis, 
$\alpha_s=-0.022^{+0.034}_{-0.032}$, even without adopting any prior on the optical depth.

\begin{figure}
\epsfig{file=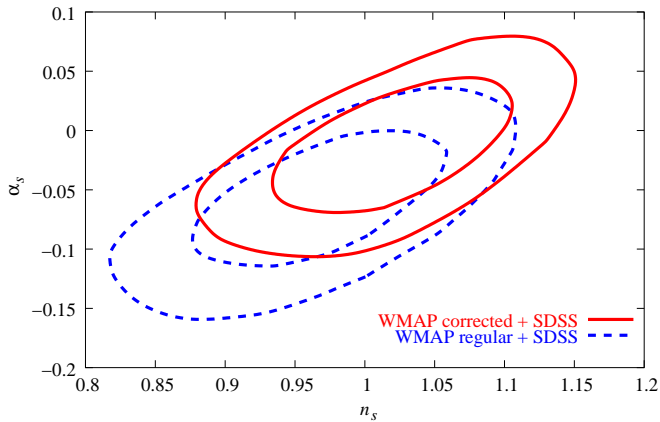, width=\linewidth}

\caption{Two dimensional contours of 68\% and 95\% probability 
in $(\alpha_s,n_s)$ plane for old 
and new MCMCs using WMAPext+SDSS data.
We use V frequency map and KP2 mask in the full likelihood analysis. 
\label{fig:runningn}
}
    
\end{figure}

As shown in table 1 the best fitted value of the primordial
slope $n_s$ increases appreciably as well, although this is mostly a
consequence of the change in running.  This is clarified in figure
\ref{fig:runningn}, which shows old and new contours in $(n_s,
\alpha_s)$ plane.  There is some degeneracy between the two
parameters, so that models with low values of running also require low
slope.
Since low values of running are excluded by the new analysis this
implies that low values of the primordial slope are also excluded,
pushing the average slope up.

\begin{table*}
\noindent
{\footnotesize Table 1: median value, $1\sigma$ and $2\sigma$
constraints on cosmological parameters for various MCMCs based on 
WMAP data alone.  5p denotes
varying 5 basic cosmological parameters in MCMCs, while 8p stands for
8 parameter chains. Old stands for the evaluation of the WMAP
likelihood using the current WMAP provided software, VKP2 is our new
exact likelihood evaluation analysis of V maps using KP2 mask and VKP0
is the same for KP0 mask.
\begin{center}
\begin{tabular}{|l|c|c|c|c|c|}
\hline
& 5p old  &5p VKP2 & 5p VKP0 & 
8p old & 8p VKP2 
\\
& & & & &  \\
\hline 
& & & & & \\
$10^2\omega_b$ & 
$2.40^{+0.06}_{-0.06}\;{}^{+0.12}_{-0.13}$ &
$2.38^{+0.06}_{-0.07}\;{}^{+0.13}_{-0.13}$ &
$2.39^{+0.06}_{-0.06}\;{}^{+0.13}_{-0.13}$ &
$2.37^{+0.17}_{-0.16}\;{}^{+0.35}_{-0.32}$ &
$2.49^{+0.19}_{-0.17}\;{}^{+0.39}_{-0.34}$

\\
& & & & &  \\
$\Omega_m$ &
$0.29^{+0.08}_{-0.06}\;{}^{+0.16}_{-0.11}$ &
$0.24^{+0.07}_{-0.05}\;{}^{+0.15}_{-0.10}$ &
$0.26^{+0.07}_{-0.06}\;{}^{+0.16}_{-0.11}$ &
$0.20^{+0.07}_{-0.06}\;{}^{+0.16}_{-0.10}$ &
$0.15^{+0.06}_{-0.04}\;{}^{+0.13}_{-0.07}$ 
\\
& & & & &  \\
$\omega_{\rm cdm}$ & 
$0.12^{+0.017}_{-0.017}\;{}^{+0.03}_{-0.03}$ &
$0.11^{+0.016}_{-0.016}\;{}^{+0.03}_{-0.03}$ &
$ 0.11^{+0.017}_{-0.016}\;{}^{+0.03}_{-0.03}$ &
$0.10^{+0.017}_{-0.017}\;{}^{+0.03}_{-0.03}$ &
$0.09^{+0.016}_{-0.015}\;{}^{+0.03}_{-0.03}$
\\
& & & & &  \\
$\tau$ & 
$0.17^{+0.04}_{-0.04}\;{}^{+0.08}_{-0.09}$ &
$0.21^{+0.04}_{-0.04}\;{}^{+0.07}_{-0.08}$ &
$0.19^{+0.04}_{-0.04}\;{}^{+0.08}_{-0.08}$&
$0.23^{+0.05}_{-0.08}\;{}^{+0.07}_{-0.16}$ &
$0.24^{+0.05}_{-0.08}\;{}^{+0.06}_{-0.17}$ 
\\
& & & & &  \\
$\sigma_8$ & 
$0.94^{+0.07}_{-0.08}\;{}^{+0.13}_{-0.17}$ &
$0.90^{+0.08}_{-0.09}\;{}^{+0.15}_{-0.19}$ &
$0.92^{+0.08}_{-0.09}\;{}^{+0.15}_{-0.19}$ &
$0.81^{+0.12}_{-0.13}\;{}^{+0.25}_{-0.26}$ &
$0.75^{+0.13}_{-0.13}\;{}^{+0.24}_{-0.25}$
\\
& & & & &  \\
$h$ & 
$0.72^{+0.05}_{-0.05}\;{}^{+0.10}_{-0.08}$ &
$0.75^{+0.05}_{-0.05}\;{}^{+0.11}_{-0.09}$ &
$0.73^{+0.05}_{-0.05}\;{}^{+0.11}_{-0.09}$&
$0.78^{+0.08}_{-0.07}\;{}^{+0.19}_{-0.13}$ &
$ 0.87^{+0.09}_{-0.08}\;{}^{+0.19}_{-0.15}$
\\
& & & & &  \\
$T/S$ & 0 & 0& 0&  $<0.76$ (95\%) & $<0.81$ (95\%) 
\\
& & & & & \\
$n_s$ &
1 & 1 & 1 
& $0.95^{+0.07}_{-0.07}\;{}^{+0.14}_{-0.15}$ &
$1.02^{+0.07}_{-0.07}\;{}^{+0.15}_{-0.15}$ 
\\
& & & & &  \\
$\alpha_s$ &
0 & 0 & 0 &
$-0.08^{+0.05}_{-0.06}\;{}^{+0.10}_{-0.13}$ &
$-0.04^{+0.05}_{-0.06}\;{}^{+0.10}_{-0.13}$ 
\\
& & & & &  \\

\hline
\end{tabular}
\end{center}
}
\label{table1}

\end{table*}

\begin{table*}
\noindent
{\footnotesize Table 2: Same as Table 1
for WMAP+SDSS (8-parameter MCMCs with regular (old) or corrected (exact 
likelihood) analysis).
The new analysis uses V KP2 with full marginalization 
and W KP2 with dust marginalization only. We also give
WMAP+SDSS+VSA (7-parameters). For the latter case we do not 
impose $\tau<0.3$. 
\begin{center}
\begin{tabular}{|l|c|c|c|c|}
\hline
& 8p SDSS+old & 8p SDSS+VKP2 & 8p SDSS+WKP2 & 7p SDSS+VSA+VKP2
\\
 & & & & \\
\hline 
 & & & & \\
$10^2\omega_b$ & 
$2.40^{+0.16}_{-0.16}\;{}^{+0.32}_{-0.30}$ &
$2.48^{+0.16}_{-0.16}\;{}^{+0.30}_{-0.31}$ &
$2.47^{+0.16}_{-0.16}\;{}^{+0.31}_{-0.30}$ &
$2.34^{+0.18}_{-0.15}\;{}^{+0.52}_{-0.28}$

\\
& & &  &\\
$\Omega_m$ &
$0.31^{+0.06}_{-0.05}\;{}^{+0.13}_{-0.08}$ &
$0.27^{+0.05}_{-0.03}\;{}^{+0.11}_{-0.06}$ &
$0.28^{+0.05}_{-0.04}\;{}^{+0.11}_{-0.07}$ &
$0.30^{+0.06}_{-0.05}\;{}^{+0.12}_{-0.10}$
\\
& & &  &\\
$\omega_{\rm cdm}$ & 
$0.128^{+0.009}_{-0.008}\;{}^{+0.019}_{-0.016}$ &
$0.121^{+0.008}_{-0.007}\;{}^{+0.017}_{-0.014}$ &
$0.123^{+0.008}_{-0.007}\;{}^{+0.017}_{-0.014}$ &
$0.123^{+0.008}_{-0.008}\;{}^{+0.017}_{-0.018}$
\\
 & & & &\\
$\tau$ & 
$0.20^{+0.07}_{-0.08}\;{}^{+0.09}_{-0.14}$ &
$0.20^{+0.07}_{-0.08}\;{}^{+0.09}_{-0.14}$ &
$0.20^{+0.07}_{-0.08}\;{}^{+0.09}_{-0.14}$ &
$0.19^{+0.11}_{-0.08}\;{}^{+0.26}_{-0.13}$ 
\\
 & & & &\\
$\sigma_8$ & 
$0.98^{+0.08}_{-0.09}\;{}^{+0.16}_{-0.16}$&
$0.97^{+0.09}_{-0.09}\;{}^{+0.16}_{-0.16}$ &
$0.97^{+0.09}_{-0.09}\;{}^{+0.16}_{-0.16}$ &
$0.93^{+0.12}_{-0.08}\;{}^{+0.29}_{-0.13}$ 
\\
 & & & &\\
$h$ & 
$0.70^{+0.05}_{-0.05}\;{}^{+0.09}_{-0.09}$ &
$0.73^{+0.04}_{-0.04}\;{}^{+0.08}_{-0.09}$ &
$0.73^{+0.04}_{-0.04}\;{}^{+0.08}_{-0.09}$ &
$0.70^{+0.05}_{-0.05}\;{}^{+0.14}_{-0.08}$
\\
& & & & \\
$T/S$ & $<0.46$ (95\%) & $<0.46$ (95\%) & $<0.47$ (95\%) & 0
\\
& & &  &\\
$n_s$ &
$0.97^{+0.06}_{-0.06}\;{}^{+0.11}_{-0.12}$ &
$1.01^{+0.05}_{-0.06}\;{}^{+0.10}_{-0.11}$ &
$1.02^{+0.05}_{-0.06}\;{}^{+0.10}_{-0.11}$ &
$0.97^{+0.06}_{-0.06}\;{}^{+0.16}_{-0.11}$ 

\\
& & & & \\
$\alpha_s$ &
$-0.060^{+0.038}_{-0.039}\;{}^{+0.074}_{-0.083}$&
$-0.015^{+0.036}_{-0.037}\;{}^{+0.072}_{-0.080}$ &
$-0.032^{+0.036}_{-0.038}\;{}^{+0.072}_{-0.080}$ &
$-0.022^{+0.034}_{-0.032}\;{}^{+0.069}_{-0.062}$
\\
& & & &\\

\hline
\end{tabular}
\end{center}
}
\label{table2}

\end{table*}

\section{Conclusions}

In this paper we have developed routines to calculate the exact
likelihood of the low resolution WMAP data.  We have projected out
unwanted foreground components by adding the foreground templates to
our covariance matrix with large variance.  Both of these methods have
not been applied to WMAP data before and should improve upon the
existing analyses.  We have tested the robustness of our results by
applying the method to many different combinations of observing frequency, 
mask, smoothing and templates and found consistent results among these various cases.
In particular, we find consistent results if we marginalize only 
over dust in $W$ channel as opposed to all 3 foreground templates, 
if we use templates 
external to WMAP instead of WMAP MEM templates,
if we use KP0 instead of KP2 mask, if we 
use ILC maps instead of individual V or W frequencies 
or if we use Healpix windows instead of gaussian smoothing. 
The two most important features of our procedure are thus 
marginalization over
dust and exact likelihood analysis. 

Important differences exist between our results and previous work. 
We find higher values of the lowest
multipoles, which is partly a consequence of template subtraction
method used in WMAP analysis. This procedure would certainly
remove some of the real power, although it is difficult to estimate
how much and the differences could also be just a statistical
fluctuation.  For the maximum likelihood value of the quadrupole we
find values between the original WMAP analysis and subsequent
reanalysis by \cite{2004MNRAS.348..885E}. The differences are within
the estimated error of the foreground contamination and we argue that
the actual value is not very reliable given how broad the likelihood
is at the peak.  More important is the shape of the likelihood
function, which we find to be broader than in the WMAP team provided
likelihood evaluation, which underestimates the errors compared to our
analysis.  This lowers the statistical significance of the departure
of the data from the concordant model. Within a Bayesian context and
assuming a flat prior on the distribution of quadrupoles we find the
probability that a model exceeds the concordance model predicted quadrupole to
be 10\%.  We also do not find anything particularly unusual in the
correlation function and in the joint quadrupole-octopole analysis. 

We combine the full likelihood calculation with foreground
marginalization at low $\ell$ with the original WMAP PCL analysis at
high $\ell$ to generate Monte Carlo Markov Chains, whose distribution
converges to the probability distribution of theoretical models given
the data and assumed priors.  The main effect of the new analysis is
on the running of the spectral index, for which the marginal 2 sigma evidence
for $\alpha_s<0$ present in the original analysis and in the recent 
analysis of WMAP+VSA \cite{2004astro.ph..2466R} 
(see also \cite{2004astro.ph..2359R}) is 
reduced to below 1 sigma.  
Using the exact WMAP likelihood analysis will be essential for 
attempts to determine the running of the spectral index by combining 
WMAP with either the 
small scale CMB data or with the upcoming Ly-$\alpha$ 
forest analysis from SDSS. In all of these cases the exact method
increases the value of the running by pushing up the CMB 
spectrum at large scales. 
Another parameter which is significantly affected is the
matter density $\Omega_m$ or, equivalently, the dark energy density
$\Omega_{\Lambda}$. We find $\Omega_m$ to be reduced by the new
analysis because of the added power at low multipoles, which is most
easily accounted for by an increase in ISW contribution. 

We have shown that the effects of the improved likelihood analysis
presented here can be significant for the determination of
cosmological parameters.  
We expect the methods applied here will be
equally important for the analysis of polarization data in WMAP,
where the foregrounds play a much more important role and where a full
likelihood analysis of joint temperature and polarization data is
necessary to extract the maximum amount of information.
Current analysis of temperature-polarization data 
is rather unsatisfactory, since it is based on the cross-spectrum 
information alone.  Without having access to the full polarization maps
we cannot improve upon it here. 
Thus the results shown in tables 1-2 
should still be viewed as preliminary
regarding the optical depth, which is 
essentially determined by the polarization data. Upcoming 
WMAP 2-year analysis/release of polarization data should elucidate 
the current situation. 
The code developed here will be made available to the community 
at \texttt{cosmas.org}.  



\section*{ACKNOWLEDGEMENTS}
We thank WMAP for the wonderful data they produced and made available through 
the LAMBDA web site. 
Our MCMC simulations were run on a Beowulf cluster at Princeton University,
supported in part by NSF grant AST-0216105.
US thanks O. Dore, C. Hirata, P. McDonald and D. Spergel for useful
discussions.  US is supported by Packard Foundation, Sloan Foundation,
NASA NAG5-1993 and NSF CAREER-0132953.

\bibliography{../Papers/BibTeX/cosmo,../Papers/BibTeX/cosmo_preprints}   

\end{document}